\documentclass[10pt]{article}

\usepackage{amsmath}
\usepackage{amssymb}
\usepackage{graphics}
\usepackage{rotating}
\usepackage{cite}
\usepackage{color}
\usepackage{fancybox}
\usepackage{pstricks}
\usepackage{xcolor}
\usepackage{multicol}
\def\0{\mbox{\tiny $0$}}
\def\1{\mbox{\tiny $1$}}
\def\2{\mbox{\tiny $2$}}
\def\3{\mbox{\tiny $3$}}
\def\4{\mbox{\tiny $4$}}
\def\5{\mbox{\tiny $5$}}
\def\6{\mbox{\tiny $6$}}
\def\7{\mbox{\tiny $7$}}
\def\8{\mbox{\tiny $8$}}
\def\9{\mbox{\tiny $9$}}
\newcommand{\cd}{black!50!red!90!}
\newcommand{\cj}{black!60!green!80!}
\newcommand{\cb}{gray!15}
\newcommand{\ct}{black!30!blue}
\begin{document}
%
%%%%%%%%%%%%%%%%%%%%%%%%%%%%%%%%%%%%%%%%%%%%%%%%%%%%%%%%%%%%%%%
\thispagestyle{empty}
\setcounter{page}{0}

{\Large \bf \color{\cj}
\begin{center}
 \fbox{\colorbox{\cb}{  \color{\ct}
\begin{tabular}{c}
TRANSVERSAL SYMMETRY BREAKING AND\\ AXIAL SPREADING MODIFICATION\\ FOR GAUSSIAN OPTICAL BEAMS
 \end{tabular}
 }}
\end{center}
}

\vspace*{1cm}

{\Large \bf
\begin{center}
\color{\cd} $\boldsymbol{\bullet}$
\color{\cj} Journal of Modern Optics 63, 417-427 (2016)
\color{\cd} $\boldsymbol{\bullet}$
\end{center}
}

\vspace*{1cm}

\begin{center}
\begin{tabular}{cc}
\begin{minipage}[t]{0,55\textwidth}
\vspace*{-0.4cm}
\color{black}
{\bf  Abstract.}\\
For a long time it was believed there was no reason to include the geometrical  phase in studying
the propagation  of gaussian optical beams through dielectric blocks.  This can be justified by the fact that the first order term in the Taylor expansion  of this phase is responsible for the lateral shift of the optical beam which is also predicted  by ray optics. From this point of view, the geometrical  phase can be seen  as a purely auxiliary concept. In this paper, we show how the second order term in the Taylor expansion  accounts for the symmetry breaking of the transversal spatial distribution and  acts  as an axial spreading modifier. These new effects clearly  shows the importance of the geometrical  phase  in describing the correct  behavior of light. To test our theoretical predictions,  we briefly discuss a  possible experimental implementation.
\end{minipage}
&
{\color{\cj} \fbox{\hspace*{-0.12cm} \color{black} {\colorbox{\cb}{
\begin{minipage}[t]{0,4\textwidth}
{\bf Manoel P. Ara\'ujo}\\
Institute of  Physics ``Gleb Wataghin''\\
State University of Campinas (Brazil)\\
{\bf \color{\ct} mparaujo@ifi.unicamp.br}
%\hrule
%\vspace*{0.15cm}
%{\bf Silv\^ania A. Carvalho }\\
%Department of Applied Mathematics\\
%State University of Campinas (Brazil)\\
%{\bf \color{\ct  silalves@ime.unicamp.br}
\hrule
\vspace*{0.15cm}
{\bf Stefano De Leo}\\
Department of Applied Mathematics\\
State University of Campinas (Brazil)\\
{\bf \color{\ct} deleo@ime.unicamp.br}
%\hrule
%\vspace*{0.15cm}
%{\bf Gabriel G. Maia}\\
%Institute of  Physics ``Gleb Wataghin''  \\
%State University of Campinas (Brazil)\\
%{\bf \color{\ct} ggm11@ifi.unicamp.br}
\hrule
\vspace*{0.15cm}
{\bf Marina Lima}\\
Department of Applied Mathematics\\
State University of Campinas (Brazil)\\
{\bf \color{\ct} marina@ime.unicamp.br}
\end{minipage}
}}}}
\end{tabular}
\end{center}

\vspace*{1cm}

\begin{center}
{\color{\ct}
{\bf
\begin{tabular}{ll}
I. & INTRODUCTION \\
II. & THE OUTGOING BEAM\\
III. & BREAKING THE TRANSVERSAL SYMMETRY\\
IV. & MODIFYING THE AXIAL SPREADING\\
V. & CONCLUSIONS AND OUTLOOKS\\
& \\
& \,[\,17 pages, 5 figures\,]
\end{tabular}
}}
\end{center}

\vspace*{3cm}

{\Large
\begin{flushright}
\color{\cd} \fbox{\hspace*{-.2cm}
\colorbox{\cb}{
\,\color{\cj}$\boldsymbol{\Sigma}$
\hspace*{-.1cm}\color{\cd}$\boldsymbol{\Delta}$
\hspace*{-.15cm}\color{\cj}$\boldsymbol{\Lambda}$
}\hspace*{-.2cm}
}
\end{flushright}
}

%%%%%%%%%%%%%%%%%%%%%%%%%%%%%%%%%%%%%%%%%%%%%%%%%%%%%%%%%%%%%%%%%%
%%%%%%%%%%%%%%%%%%%%%%%%%%%%%%%%%%%%%%%%%%%%%%%%%%%%%%%%%%%%%%%%%%%%%%%

\newpage

%%%%%%%%%%%%%%%%%%%%%%%%%%%%%%%%%%%%%%%%%%%%%%%%%%%%%%%%%%%%%%%%%%%%%%%%%%
%%%%%%%%%%%%%%%%%%%%%%%%%%%%%%  SECTION I    %%%%%%%%%%%%%%%%%%%%%%%%%%%%%
%%%%%%%%%%%%%%%%%%%%%%%%%%%%%%%%%%%%%%%%%%%%%%%%%%%%%%%%%%%%%%%%%%%%%%%%%%

\section*{\large \color{\ct} I. INTRODUCTION}

 The well known analogy between non relativistic quantum mechanics\cite{Cohen1977} and
 optics\cite{Wolf1999, Saleh2007}, based on replacing quantum mechanical potentials and wave functions
 by dielectric structures and electric field amplitudes\cite{AJP1976,Yasumoto1983,AJP1988,Zanella2003}, allows to reproduce and test quantum mechanical phenomena, probabilistic in nature and valid for massive particle wave packets, by optical experiments\cite{PR07,JP2009}. The electric field is stationary and the analogy with quantum mechanics is fully understood when time in quantum mechanics is replaced by the axial spatial variable in optics\cite{LPR309}. A fascinating example of this analogy is given by  the Goos-H\"anchen shift \cite{1947AP436} which represents the optical counterpart of the non relativistic delay time\cite{vanDijk1992}. The mathematical connection between the Maxwell equations for the light propagation in the presence of a dielectric medium and  the Scr\"odinger  equation for the electron propagation in the presence of a potential step is clear when we compare  the Fresnel coefficients at the dielectric interface  with the reflection and transmission coefficients of the potential step\cite{DeLeo2008,Use2011a,Use2011b,ana1,ana2,Garrido2011,Zhu2012,Selmke2013,Car2013a,Car2013b,Car2014}.

The Fresnel coefficients do {\em not}\, usually  contain the {\em geometrical  phase}\cite{AJP}. Indeed,
due to the fact that the optical geometrical path can be directly obtained by the Snell law of ray optics,  the geometrical  phase is often neglected in the study of light propagation through dielectric blocks. What we aim to show in this paper is that whereas the first order term in the geometrical phase expansion can be  easily replaced by the ray optics results, the second order term is responsible for {\em new} interesting effects such as the {\em breaking} of symmetry in the transversal spatial distribution and  the {\em modification} of the axial spreading which {\em cannot} be predicted by ray optics and, consequently, clearly represent an evidence of the importance of the Snell optical phase in studying the quantum behavior of light.

The geometrical phase has {\em not}\, to be confused with the Goos-H\"anchen phase\cite{Tamir1971, Tamir1986,rev1,rev2,rev3} which, for incidence angles greater than critical angle, directly comes from the complex nature of the Fresnel coefficients. The first order contribution of this additional phase gives the well known  Goos-H\"anchen lateral beam displacement\cite{Manoel2013}. A study of the second order contribution is reported in ref.\cite{Carniglia1977}. To avoid confusion between the geometrical phase and the  Goos-H\"anchen  phase, we shall refer to the geometrical phase as the Snell optical phase.

As we shall see in detail later, the transversal breaking of symmetry and the axial spreading dependence on the incidence angle take their main contribution from the Snell optical phase. This is due to the fact that the Snell phase contributions are proportional to the block dimensions whereas the Goos-H\"anchen phase contributions are proportional to the beam wavelength. Consequently, as a first approximation, the Goos-H\"anchen phase contributions to the transversal breaking of symmetry and the axial spreading modification can be neglected.

The analytical expression of the second order term  of the Snell optical phase expansion allows to determine in which incidence conditions we can modify the axial spreading and  in which angle range and for which refractive index we can maximize the breaking of the transversal symmetry.

 We start our analysis by considering a gaussian laser with a fixed frequency $\omega\,=\,|\boldsymbol{k}|\,c=2\,\pi\, c/\lambda$.  Localized  optical   beams are obtained convoluting plane waves by appropriate  wave number distributions. In this paper, we use the distribution
\begin{equation}
g(k_x,\,k_y)= \exp \left[-\left(k_{x}^{^2}\,+\,k_{y}^{^2}\right)\,\mbox{w}_{\0}^{^2} \,/\, 4\right]\,\,,
\end{equation}
where $\mbox{w}_{\0}$ represents the minimal beam-waist of the gaussian laser. Consequently, the intensity of our beam, $\mbox{I}(\boldsymbol{r})=\left|E(\boldsymbol{r})\right|^{^2}$, is given by
\begin{eqnarray}
\mbox{I}(\boldsymbol{r})= \mbox{I}_{_0}\,\left| \, \frac{\mbox{w}_{\0}^{\2}}{4 \pi}\, \int^{^{+\infty}}_{_{-\infty}}\hspace*{-.5cm}\mbox{d}k_x\, \int^{^{+\infty}}_{_{-\infty}}\hspace*{-.5cm} \mbox{d}k_y\,\,g(k_x,\,k_y)\, \exp \left[\,i\,\left(\boldsymbol{k} \cdot \boldsymbol{r}\right)\right]\right|^{^2}\,\,,
\label{eq:Ein1}
\end{eqnarray}
where $k_z=\sqrt{k^{^2}-k_x^{^2}-k_y^{^2}}$ and the normalization $\mbox{w}_{\0}^{\2}/4 \pi$ has been chosen to guarantee
\begin{eqnarray}
\int^{^{+\infty}}_{_{-\infty}}\hspace*{-.5cm}\mbox{d}x\, \int^{^{+\infty}}_{_{-\infty}}\hspace*{-.5cm} \mbox{d}y\,\, \mbox{I}(\boldsymbol{r})= \pi\,\mbox{I}_{_0}/2\,\,.
\label{eq:Pot}
\end{eqnarray}
For $\mbox{w}_{\0} \gtrsim\lambda$, the beam divergence is relatively small and we can use the
paraxial approximation\cite{Saleh2007},
\begin{eqnarray}
k_z \approx k-\frac{k_x^{^2}+k_y^{^2}}{2k}\,\,.
\label{eq:Ein12}
\end{eqnarray}
This allows  to separate the $x$ and $y$ integrals and after simple algebraic manipulations to obtain
\begin{eqnarray}
\label{int}
\mbox{I}(\boldsymbol{r})\,=\,\mbox{I}_{_0} \, \mathcal{G}(x,\,z) \,\,
\mathcal{G}(y,\,z)\,\,,
\end{eqnarray}
where
\begin{eqnarray}
\mathcal{G}(x,\,z) \,=\,\frac{\mbox{w}_{\0}}{\mbox{w}(z)}\,\exp\left[- \,2\,\,\frac{x^{^2}}{\mbox{w}^{^2}(z)} \right]\,\,\,\,\,\,\,\mbox{and}\,\,\,\,\,\,\, \mbox{w}(z)\,=\,\mbox{w}_{_0}\sqrt{1 +\left(\frac{2\,z}{k\,\mbox{w}^{^2}_{\0}}\right)^{^2}}\,\,.
\end{eqnarray}
Eq.(\ref{int}) clearly shows the dependence of the optical beam spreading  on the axial coordinate $z$
and  the symmetry  between  the transversal coordinates  $x$ and $y$.

To get clear for the reader the objective of our investigation, it is interesting to briefly discuss the mathematical idea which stimulated our work. To do it, let us consider the wave number distribution
which determines the behavior of  the transmitted beam, i.e.
\begin{equation}
\label{gtra}
g_{_{T}}(k_x,\,k_y)=  T(k_x,\,k_y)\,g(k_x,\,k_y) = |T(k_x,\,k_y)|\,g(k_x,\,k_y)\,\exp[\,i\,\Phi(k_x,\,k_y)\,]\,\,,
\end{equation}
where $T(k_x,\,k_y)$ is the transmission coefficient obtained by solving the electromagnetic wave equations in the presence of stratified media. Such a transmission coefficient will be explicitly calculated in the next section. To understand the purpose of our analysis, it is sufficient to expand up to second order the optical phase which appears in Eq.(\ref{gtra}),
\begin{equation}
\Phi(k_x,\,k_y)\approx \Phi(0,0) +
\left[\frac{\partial \Phi}{\partial k_x} \right]_{_{(0,0)}}\hspace{-0.4cm}k_x +
\left[\frac{\partial \Phi}{\partial k_y} \right]_{_{(0,0)}}\hspace{-0.4cm}k_y +
\left[\frac{\partial^{^2} \Phi}{\partial k_x^{^2}} \right]_{_{(0,0)}}\hspace{-0.4cm}\frac{k_x^{^2}}{2} + \left[\frac{\partial^{^2} \Phi}{\partial k_y^{^2}} \right]_{_{(0,0)}}\hspace{-0.4cm}\frac{k_y^{^2}}{2} + \left[\frac{\partial^{^2} \Phi}{\partial k_x\partial k_y} \right]_{_{(0,0)}}\hspace{-0.4cm}k_xk_y\,\,.
\label{optphase}
\end{equation}
The first order terms are clearly responsible for the transversal lateral displacements. Indeed, they directly act on the transversal spatial phase $\exp[\,i\,(\,k_x\, x +k_y\, y\,)\,]$ modifying the center of the beam,
\begin{equation}
\{\,x\,,\,y\,\}=\{\,0\,,\,0\,\}\,\,\,\,\,\,\,\rightarrow\,\,\,\,\,\,\,\left\{\,-\,\frac{\partial \Phi}{\partial k_x}\,\,,\,  -\,\frac{\partial \Phi}{\partial k_y}   \,\right\}_{_{(0,0)}}\,\,.
\end{equation}
 The {\em pure} second order terms act on the axial spatial phase $\exp[\,-\,i\,(\,k_x^{^{2}} +k_y^{^{2}}\,)\,z/\,2\,k\,]$ modifying the axial spreading in the transversal plane $xz$ and $yz$
 \begin{equation}
  \left\{\,\mbox{w}\left(z -\,k\,  \frac{\partial^{^2} \Phi}{\partial k_x^{^2}}\right)\,\,,\,
 \mbox{w}\left(z -\,k\,  \frac{\partial^{^2} \Phi}{\partial k_y^{^2}}\right)
 \,\right\}_{_{(0,0)}}\,\,.
 \end{equation}
 The aim of our investigation is to study the possibility to  break the transversal symmetry of the beam,
\begin{equation}
\left[\frac{\partial^{^2} \Phi}{\partial k_x^{^2}} \right]_{(0,0)} \neq   \,\,\,\,\, \left[\frac{\partial^{^2} \Phi}{\partial k_y^{^2}} \right]_{(0,0)}\,\,,
\end{equation}
and, if this really happens, to analyze for which incidence angles we can maximize such a breaking of symmetry. In addition, it is also interesting to study the sign of these contributions. Positive and negative values of the pure second order terms characterize different  axial spreading behaviors. We finally observe that  the {\em mixed} second order term has {\em not} a spatial counterpart.

 This paper is organized as follows. In section II, we gives the Fresnel coefficients for the dielectric  block whose upper front view is illustrated in Fig.\,1, discuss some geometrical properties of the dielectric block, and find an analytical approximation for the intensity of the outgoing beam. In section III, we analyze  the optical Snell phase and explicitly calculate the Taylor expansion up to second order. Starting from this expansion, we discuss in which incidence angle region we can maximize the breaking of symmetry of the transversal components (section III) and see axial spreading modifications (section IV). Our conclusions, forthcoming studies, and experimental proposals are drawn in the final section.

\section*{\large \color{\ct} II. THE OUTGOING BEAM}

To calculate the Fresnel coefficients for the optical beam propagating trough an elongated prism composed by $N$ dielectric blocks, we can use  the analogy between optics and quantum mechanics. For dielectric sides very large with respect to  the beam-waist,  the transmission and reflection amplitudes  can be easily obtained by  successive applications of the step analysis at the dielectric discontinuities\cite{ManoelPRA}.

The incoming beam,
\begin{equation}
E_{_{\rm inc}}(x,y,z)= E_{_0}\, \frac{\mbox{w}_{\0}^{\2}}{4 \pi}\, \int^{^{+\infty}}_{_{-\infty}}\hspace*{-.5cm}\mbox{d}k_x\, \int^{^{+\infty}}_{_{-\infty}}\hspace*{-.5cm} \mbox{d}k_y\,\,g(k_x,\,k_y)\, \exp \left[\,i\,\left(k_xx+k_yy+k_zz\right)\right]\,\,,
\end{equation}
has its source in $S$ and moves along the $z$ axis, see Fig.\,1(a).
 Due to the fact that the first air/dielectric discontinuity is along the $\tilde{z}$-axis, it is
convenient to rewrite the Maxwell equations  by using the coordinate system $(x,\tilde{y},\tilde{z})$,
\begin{equation}
\left[\, \partial_{xx} + \partial_{\tilde{y}\tilde{y}} +
\partial_{\tilde{z}\tilde{z}} - n^{2}(\tilde{z})\,k^{^{2}}\,\right]\,
E_{_{\rm left}} (x,\tilde{y},\tilde{z}) =
0\,,\,\,\,\,\,\,\,\,\,\,\mbox{with}\,\,\,n(\tilde{z})=\left\{\begin{array}{l}
1\,\,\,\mbox{for}\,\,\,
\tilde{z}<\overline{S\widetilde{D}}\,\,,\\
n\,\,\,\mbox{for}\,\,\,
\tilde{z}>\overline{S\widetilde{D}}\,\,.
\end{array}
\right.
\label{MEleft}
\end{equation}
 The plane wave solutions are then given by
\begin{equation}
\label{eqleft}
\exp\left[\,i\, (\,k_{x} x + k_{\tilde{y}}\, \tilde{y} \,)\,\right]
\,\times \,\begin{cases}
\exp\left[\,i\,k_{\tilde{z}}\, \tilde{z}\,\right] + R_{_{\rm
left}}\,\exp\left[\,-\,i\,k_{\tilde{z}}\, \tilde{z}\,\right] &
\mbox{for}\,\,\,
\tilde{z}<\overline{S\widetilde{D}}\,\,,\\
T_{_{\rm left}}\,\exp\left[\,i\,q_{\tilde{z}}\, \tilde{z}\,\right] &
 \mbox{for}\,\,\,
\tilde{z}>\overline{S\widetilde{D}}\,\,,
\end{cases}
\end{equation}
 where
\begin{equation}
\left(\, k_{\widetilde{y}}\,,\,  k_{\widetilde{z}}\,\right)=  \left(\,k_y\,\cos\,\theta +k_z\,\sin \theta\,,\, -\,k_y\,\sin\,\theta +k_z\,\cos\, \theta \,\right)\,\,\,\,\,\,\,\mbox{and}\,\,\,\,\,\,\,
q_{\widetilde{z}} = \sqrt{n^{^2}k^{^{2}} - k_{x}^{^2} - k_{\widetilde{y}}^{^2}}\,\,.
\label{eq:qyqz}
\end{equation}
Observe that the component parallel to the discontinuity is not changed passing from air to dielectric. By using the quantum mechanics step analysis and imposing continuity at $\tilde{z}=\overline{S\widetilde{D}}$ for the solution and its $\tilde{z}$-derivative,  we can immediately find the transmission coefficient at the left interface, i.e.
\begin{equation}
T_{_{\rm left}} = \displaystyle{\frac{2\, k_{\widetilde{z}}}{k_{\widetilde{z}} + q_{\widetilde{z}}}}\,\, \exp\left[\,i \,( k_{\widetilde{z}} - q_{\widetilde{z}} )\, \overline{S\widetilde{D}}\, \right]\,\,.
\end{equation}
To obtain the reflected field at the down interface, we have to introduce
the system of coordinates $y_*$ and $z_*$, where $z_*$  is the axis perpendicular to the up and down  interfaces, see Fig.\,1(a), and solve the following equation
\begin{equation}
\left[\, \partial_{xx} + \partial_{y_{_{*}}y_{_{*}}} +
\partial_{z_{_{*}}z_{_{*}}} - n^{2}(\tilde{z})\,k^{^{2}}\,\right]\,
E_{_{\rm down}} (x,y_{_{*}},z_{_{*}}) =
0\,,\,\,\,\,\,\,\,\,\,\,\mbox{with}\,\,\,n(z_{_{*}})=\left\{\begin{array}{l}
n\,\,\,\mbox{for}\,\,\,
z_{_{*}}<\overline{SD_{_*}}\,\,,\\
1\,\,\,\mbox{for}\,\,\,
z_{_{*}}>\overline{SD_{_{*}}}\,\,.
\end{array}
\right.
\label{MEdown}
\end{equation}
In this case, the plane wave solutions are
\begin{equation}
\label{eqleft}
\exp\left[\,i\, (\,k_{x} x + q_{y_{_{*}}}\, y_{_{*}} \,)\,\right]
\,\times \,\begin{cases}
\exp\left[\,i\,q_{z_{_{*}}}\, \tilde{z}\,\right] + R_{_{\rm
down}}\,\exp\left[\,-\,i\,q_{z_{_{*}}}\, z_{_{*}}\,\right] &
\mbox{for}\,\,\,
z_{_{*}}<\overline{SD_{_{*}}}\,\,,\\
T_{_{\rm down}}\,\exp\left[\,i\,k_{z_{_{*}}}\, z_{_{*}}\,\right] &
 \mbox{for}\,\,\,
z_{_{*}}>\overline{SD_{_{*}}}\,\,,
\end{cases}
\end{equation}
 where
\begin{equation}
 \left(\,q_{y_{_{*}}}\,,\,q_{z_{_{*}}}\,\right) =
 \left(\,\frac{k_{\widetilde{y}}\,+\,q_{\widetilde{z}}}{\sqrt{2}}\,,\, \frac{-\,k_{\widetilde{y}}\,+\,q_{\widetilde{z}}}{\sqrt{2}}\,\right)
 \,\,\,\,\,\,\,\mbox{and}\,\,\,\,\,\,\,
 k_{z_{_{*}}} = \sqrt{k^{^{2}} - k_{x}^{^2} - q_{y_{_{*}}}^{^2}}\,\,.
\end{equation}
By imposing continuity at $z_{_{*}}=\overline{SD_{_{*}}}$ for the solution and its $z_{_{*}}$-derivative,  we then find the reflection coefficient at the down interface,
\begin{equation}
R_{_{\rm down}} =
 \frac{q_{z_{_{*}}} - k_{z_{_{*}}}}{q_{z_{_{*}}} + k_{z_{_{*}}}} \,\,\exp\left[\,2 \,i \,q_{z_{_{*}}}\, \overline{S D_{_{*}}}\,\right]\,\,.
\end{equation}
The reflection coefficient at the up interface can be directly obtained  from $R_{_{\rm down}}$
by the following substitutions
\[  (\,k_{z_{_{*}}}\,,\,q_{z_{_{*}}}\,)\,\,\,\rightarrow\,\,\,  -\,\, (\,k_{z_{_{*}}}\,,\,q_{z_{_{*}}}\,)  \]
and observing that the discontinuity is now located at
\[ z_{_{*}} = -\, \,  \left(\frac{\overline{AB}}{\sqrt{2}} - \overline{S D_{_{*}}}\right)\,\,. \]
Consequently, we find
\begin{equation}
R_{_{\rm up}} =
 \frac{q_{z_{_{*}}} - k_{z_{_{*}}}}{q_{z_{_{*}}} + k_{z_{_{*}}}} \,\,\exp\left[\, 2 \,i \,q_{z_{_{*}}}\, \left(\frac{\overline{AB}}{\sqrt{2}} - \overline{S D_{_{*}}}\right)\,\right] \,\,.
\end{equation}
Finally, the last transmission coefficient (right interface) is obtained from $T_{_{\rm left}}$  by interchanging
\[  k_{\widetilde{z}} \,\,\,\leftrightarrow\,\,\, q_{\widetilde{z}}\]
and using the fact that the right discontinuity is located at
\[ \tilde{z} = \overline{S\widetilde{D}}+\frac{\overline{BC}}{\sqrt{2}}\,\,.\]
This leads to
\begin{equation}
T_{_{\rm right}} = \frac{2\, q_{\widetilde{z}}}{q_{\widetilde{z}} + k_{\widetilde{z}}} \,\, \exp\left[\,i \,( q_{\widetilde{z}} -k_{\widetilde{z}} )\, \left(\,\overline{S\widetilde{D}}+\frac{\overline{BC}}{\sqrt{2}}\,\right) \right]\,\,.
\end{equation}
These coefficients represent the Fresnel coefficients for $s$-polarized waves\cite{DeLeo2008,Use2011a,Use2011b}. To obtain the coefficients  for $p$-polarization, we can use the following translation rules\cite{Car2013a,Car2013b}
\begin{equation}
(\,k_{z_{_{*}}},\,k_{\widetilde{z}}\,)\,\longrightarrow\,n\,(\,k_{z_{_{*}}},\,k_{\widetilde{z}}\,)\,\,, \qquad (\,q_{z_{_{*}}},\,q_{\widetilde{z}}\,)\,\longrightarrow\,(\,q_{z_{_{*}}},\,q_{\widetilde{z}}\,)\,/\,n
\,\,.
\end{equation}
It has to be observed here that the previous translation rules  do not apply to the geometrical phase which appears in the reflection and transmission coefficients.  Finally, the transmission coefficients for $s$-$p$ polarized light which propagates through  an elongated prism composed by $N$ dielectric blocks are given by
\begin{eqnarray}
\displaystyle{T^{^{[s,p]}}(k_{x},k_{y}) = \left\{\frac{4\, q_{\widetilde{z}}\, k_{\widetilde{z}}}{(q_{\widetilde{z}} + k_{\widetilde{z}})^{\2}} \left(\frac{q_{z_{_{*}}} - k_{z_{_{*}}}}{q_{z_{_{*}}} + k_{z_{_{*}}}}\right)^{^{2N}}\hspace{-0.2cm}, \,\,\,\frac{4\, n^{\2}\, q_{\widetilde{z}}\, k_{\widetilde{z}}}{(q_{\widetilde{z}} + n^{\2}\, k_{\widetilde{z}})^{\2}} \left(\frac{q_{z_{_{*}}} - n^{\2}\, k_{z_{_{*}}}}{q_{z_{_{*}}} + n^{\2}\, k_{z_{_{*}}}}\right)^{^{2N}}\right\} \,\exp[\,i\, \phi_{_{\rm Snell}}\,]}\,\,,
\label{Tsp}
\end{eqnarray}
where
\begin{align}
\phi_{_{\rm Snell}}\,&=\,N\,\left[\sqrt{2} \,q_{z_{_{*}}}\, \overline{AB}+\, \left(\,q_{\widetilde{z}} - k_{\widetilde{z}}\,\right)\,\frac{\overline{BC}}{\sqrt{2}}\,\right]  \,\,,
\label{eq:snell}
\end{align}
represent the geometric phase. As we shall show in detail in the next section, the Snell phase (\ref{eq:snell}) allows, by using the stationary phase method\cite{Artmann1948,SPM1,SPM2}, to obtain  the Snell law of ray optics\cite{AJP}. For total internal reflection, $\theta>\theta_{_{\rm c}}$, the reflection coefficients at the up and down dielectric/air interfaces becomes complex, $k_{z_*}^2<0$, and we get an additional phase. This additional phase,
\begin{equation}
\phi^{^{[s]}}_{_{\rm GH}} = -4N\arctan\left[\,|k_{z_*}|\,/\,q_{z_*}\right]\,, \qquad \phi^{^{[p]}}_{_{\rm GH}} = -4N\arctan\left[\,n^{^2}|k_{z_*}|\,/\,q_{z_*}\right]\,\,,
\label{eq:GHphase}
\end{equation}
is responsible for a new additional lateral shift, known as Goos-H\"anchen shift. This shift was experimentally observed in 1947\cite{1947AP436} and, one year later, explained by  Artmann  which proposed an analytical expression\cite{Artmann1948}. For a detailed discussion of the additional lateral shift, we refer the reader to\cite{rev1,rev2,rev3,Manoel2013}.

The total optical phase thus contains two terms,
\begin{equation}
\displaystyle{\Phi^{^{[s,\,p]}} = \phi_{_{\rm Snell}}+\phi^{^{[s,p]}}_{_{\rm GH}}}\,\,.
\end{equation}
The first one, the Snell term, is independent of  polarization and contains the geometrical optical path predicted by the Snell law. The second one, the Goos-H\"anchen term, is polarization dependent and only appears  for total internal reflection.

Observing that the reflection and transmission coefficients are  quadratic in $k_x$, we immediately get
\begin{equation}
\left[\frac{\partial \Phi^{^{[s,\,p]}}}{\partial k_x} \right]_{(0,0)} =\,\,\,\, \left[\frac{\partial^{^2} \Phi^{^{[s,\,p]}}}{\partial k_x \partial k_y} \right]_{(0,0)}=\,\,\,\, 0\,\,,
\end{equation}
and consequently, by using the Taylor expansion of the optical phase, we can rewrite  the transmission coefficient as follows
\begin{equation}
\displaystyle{T^{^{[s,p]}}(k_x,\,k_y) \,\approx\, |T^{^{[s,p]}}(0,\,0)|\,\exp\left[i\left(\,  \Phi(0,0)+ \left[\frac{\partial \Phi}{\partial k_y} \right]_{(0,0)}\,\hspace{-0.4cm}k_y + \left[\frac{\partial^{^2} \Phi}{\partial k_x^{^2}} \right]_{(0,0)}\hspace{-0.4cm}\frac{k_x^{^2}}{2} + \left[\frac{\partial^{^2} \Phi}{\partial k_y^{^2}} \right]_{(0,0)}\hspace{-0.4cm}\frac{k_y^{^2}}{2}\right)\right]}\,,
\label{eq:trans_approx}
\end{equation}
where
\[
T^{^{[s]}}(0,0) =
\frac{4\,n\,\cos\psi\,\cos\theta}{(\cos\theta\,+\,n\cos\psi)^{^2}}
\left(\frac{n\,\cos\varphi\,-\,\sqrt{1\,-\,n^{^2}\sin^{^2}\varphi}}{n\,
\cos\varphi\,+\,\sqrt{1\,-\,n^{^2}\sin^{^2}\varphi}}\right)^{^{2N}}
\]
and
\[T^{^{[p]}}(0,0)=
\frac{4\,n\,\cos\psi\,\cos\theta}{(n\,\cos\theta\,+\,\cos\psi)^{^2}}
\left(\frac{\cos\varphi\,-\,n\,\sqrt{1\,-\,n^{^2}\sin^{^2}\varphi}}{\cos\varphi\,+
\,n\,\sqrt{1\,-\,n^{^2}\sin^{^2}\varphi}}\right)^{^{2N}}
\]
are the well known Fresnel coefficients for plane waves\cite{Wolf1999,Saleh2007}.

By using the approximation (\ref{eq:trans_approx}), the  integrals  which appear in the  transmission intensity
\begin{equation}
\mbox{I}_{_T}^{^{[s,\,p]}}(\boldsymbol{r}) = \mbox{I}_{_0}\,\left|\frac{\mbox{w}_{\0}^{^2}}{4 \pi} \int^{^{+\infty}}_{_{-\infty}}\hspace*{-.5cm}\mbox{d}k_{x} \int^{^{+\infty}}_{_{-\infty}}\hspace*{-.5cm}\mbox{d}k_{y}\,\, T^{^{[s,\,p]}}\hspace{-0.1cm}(k_x,\,k_{y}) \, g(k_x,\,k_y)\,e^{ikz} \exp \left[i\left(k_x\,x + k_y\,y -\frac{k_x^{^2}+k_y^{^2}}{2k}\,z\right)\right] \right|^{^2}
\label{eq:Eout1}
\end{equation}
can be analytically solved leading to
\begin{equation}
\mbox{I}_{_T}^{^{[s,\,p]}}(\boldsymbol{r})= \mbox{I}_{_0}\,|T^{^{[s,\,p]}}(0,\,0)|^{^2}\,\mathcal{G}\left(x,\,\,z-k\,\left[\frac{\partial^{^2} \Phi}{\partial k_x^{^2}} \right]_{(0,0)} \right)\,\mathcal{G}\left(y+\left[\frac{\partial \Phi}{\partial k_y} \right]_{(0,0)}\hspace*{-0.2cm},\,\,z -k\,\left[\frac{\partial^{^2} \Phi}{\partial k_y^{^2}} \right]_{(0,0)} \right) \,\,.
\label{eq:Eout2}
\end{equation}
It is clear from the previous expression that the first order contribution of the optical phase acts as an $y$ translation operator generating the Snell and Goos-H\"anchen lateral displacements, whereas the second order term, modifying the axial spreading, break the transversal symmetry. Due to the fact that
 \begin{equation}
\left\{\,\frac{\partial \phi_{_{\rm GH}}}{\partial k_y}  \,\mbox{\huge $/$}\,\frac{\partial \phi_{_{\rm Snell}}}{\partial k_y}\,,\,
\frac{\partial^{^2} \phi_{_{\rm GH}}}{\partial k_y^{^2}}  \,\mbox{\huge $/$}\,\frac{\partial^{^2} \phi_{_{\rm Snell}}}{\partial k_y^{^2}}\,,\,
\frac{\partial^{^2} \phi_{_{\rm GH}}}{\partial k_x^{^2}}  \,\mbox{\huge $/$}\,\frac{\partial^{^2} \phi_{_{\rm Snell}}}{\partial k_x^{^2}}
\,
 \right\}_{(0,0)} \propto\,\,\,\,\,  \frac{\lambda}{\overline{AB}}\,\left\{\,1\,,\,1\,,\,1\,\right\}\,\,,
 \end{equation}
 to study the breaking of symmetry and axial spreading effects we can use, without loss of generality, the following approximated transmission intensity
 \begin{equation}
\mbox{I}_{_T}^{^{[s,\,p]}}(\boldsymbol{r})= \mbox{I}_{_0}\,\mbox{P}^{^{[s,\,p]}}_{_{T}}\,\mathcal{G}(x,\,\,z-z_{_{\rm Snell}}^{^{\perp}})\,\mathcal{G}\left(y-y_{_{\rm Snell}},\, z-z_{_{\rm Snell}}   \right) \,\,,
\label{eq:Eout3}
 \end{equation}
where $\mbox{P}^{^{[s,\,p]}}_{_{T}}=|T^{^{[s,\,p]}}(0,\,0)|^{^2}$ and
\begin{equation}
\left\{\,y_{_{\rm Snell}}\,,\, z_{_{\rm Snell}}\,,\, z_{_{\rm Snell}}^{^{\perp}}\,\right\} =
\left\{\,-\,\frac{\partial \phi_{_{\rm Snell}}}{\partial k_y}\,,\,k\,
\frac{\partial^{^2} \phi_{_{\rm Snell}}}{\partial k_y^{^2}}\,,\,k\,
\frac{\partial^{^2} \phi_{_{\rm Snell}}}{\partial k_x^{^2}}
\,
 \right\}_{(0,0)}\,\,.
 \label{defin}
\end{equation}

\section*{\large \color{\ct} III. BREAKING THE TRANSVERSAL SYMMETRY}

To build an elongated prism which guarantees $2\,N$ internal reflections, we have to impose the following geometrical constraint between the sides $\overline{AB}$ and $\overline{BC}$ of the single dielectric  block\cite{ManoelPRA}
\begin{equation}
\overline{BC} =\sqrt{2} \,\tan \varphi \,\, \overline{AB}\,\,,
 \label{eq:c}
\end{equation}
with $\varphi=\psi+\frac{\pi}{4}$ and $\psi$ obtained from the incidence angle $\theta$  by using the Snell law, $\sin\theta = n\sin\psi$, see Fig.\,1(a). In terms of the incidence angle $\theta$,   the previous constraint becomes
\begin{equation}
\overline{BC} = \sqrt{2}\,\,\,\frac{n^{\2} + 2 \, \sin\theta\,\sqrt{n^{^2}-\sin^{\2}\theta}}{n^{^{2}}-2\,\sin^{\2}\theta}\,
\, \,\,\overline{AB}\,\,.
 \label{cons}
\end{equation}
For convenience of presentation and for geometrical constraints, we fix to $[\,-\,\pi/6\,,\,\pi/4\,]$ the range of the incidence angle. Consequently, the block dimension varies from
\begin{equation}
\overline{BC}_{_{-\,\pi/6}} = \sqrt{2}\,\,\,\frac{2\,n^{\2} -  \sqrt{4\,n^{^2}-1}}{2\,n^{^{2}}-1}\,
\,\, \,\overline{AB}
\,\,\,\,\,\,\,\,\,\,\mbox{to}\,\,\,\,\,\,\,\,\,\,
\overline{BC}_{_{\pi/4}} = \sqrt{2}\,\,\,\frac{n^{\2} +  \sqrt{2\,n^{^2}-1}}{n^{^{2}}-1}\,
\,\,\, \overline{AB}\,\,.
 \label{eq:c3}
\end{equation}
For $n=\sqrt{2}$ this means, see Fig.\,1(b,d)
\begin{equation}
\overline{BC}_{_{n=\sqrt{2}}} \,\,\in\,\,\sqrt{2}\,\left[\,\frac{4-\sqrt{7}}{3}\,\,,\,\,2+\sqrt{3}\,\right]\,\,\overline{AB}\,\,\approx\,\,
\left[\,0.638\,\,,\,\,5.278\,\right]\,\,\overline{AB}
\,\,.
\end{equation}
By increasing the value of the refractive index we decrease the previous interval. For example for
BK7/LASF9 glasses and Diamond which for an incidence optical beam with $\lambda=633\,{\rm mm}$ [He-Ne laser]
have the following refractive index
\[
n_{_{\rm BK7}}=1.515\,\,,\,\,\,\,\,n_{_{\rm LASF9}}=1.845\,\,\,\,\,\,\,\mbox{and}\,\,\,\,\,\,\,
n_{_{\rm Diamond}}=2.412\,\,,\]
we find
\begin{equation}
\overline{BC}\,/\,\overline{AB}\,\,\,\,\in\,\,\,\,
\left[\,0.682\,,\,4.574\,\right]_{_{\rm BK7}}\,\,,\,\,\,\,\,\,
\left[\,0.793\,,\,3.420\,\right]_{_{\rm LASF9}}\,\,,\,\,\,\,\,
\left[\,0.920\,,\,2.665\,\right]_{_{\rm Diamond}}\,\,.
\end{equation}

Let us start by considering  the first order contribution of the Snell phase. From Eq.\,(\ref{eq:snell}),  by using Eq.\,(\ref{eq:c}) and after simple algebraic manipulations, we obtain
\begin{equation}
y_{_{\rm Snell}}= N\,\,\frac{\cos\theta -\sin \theta}{\sqrt{2}}\, \, \overline{BC}\,\,,
\end{equation}
which  represents the lateral shift predicted by the Snell law of ray optics. For example for a single block, we find
\begin{equation}
\frac{y_{_{\rm Snell}}}{\overline{BC}} = \left\{\begin{array}{lcll}
(\sqrt{3}+1)\,/\,2\,\sqrt{2} &\,\,\,& [\,\theta=-\pi/6\,, &\mbox{Fig.\,1(b)}\,]\,\,,\\
1\,/\,\sqrt{2} & &  [\,\theta=0\,,& \mbox{Fig.\,1(c)\,}\,]\,\,,\\
0 & &  [\,\theta=\pi/4\,,&\mbox{Fig.\,1(d)}\,]\,\,.
\end{array} \right.
\end{equation}
Observe that the dependence on the refractive index $n$ is contained in the $\overline{BC}$ side dimension, see Eq.\,(\ref{cons}).

As discussed in the previous section, the second order contributions of the Snell phase introduce
 transversal shifts in the axial coordinate $z$,
\begin{equation}
\label{eqzout}
z \geq z_{_{\rm out}} = \overline{SD} + N\, \frac{\cos\theta +\sin \theta}{\sqrt{2}}\, \, \overline{BC} = \overline{SD} + N\, (\,\cos\theta +\sin \theta\,)\,\tan \varphi \,\overline{AB} \,\,,
\end{equation}
and break the symmetry in the transversal components of the gaussian laser.  These transversal shifts can be explicitly  found by calculating the second derivative of the Snell phase  with respect to $k_x$ and $k_y$.
By using Eqs.\,(\ref{eq:snell}) and (\ref{defin}), we find
\begin{eqnarray}
\label{eqzsnell1}
z_{_{\rm Snell}}^{^{\perp}}&= & N\, \left[\sin\theta\,+\,\cos\theta\,\tan\varphi\,-\,\frac{\cos^{^2}\theta}{n\,\cos\psi}\,
(1\,+\,\tan\varphi)\,\right]\,\overline{AB}\,\,,\nonumber \\
z_{_{\rm Snell}}& = & N\,
\left[(\sin\theta\,+\,\cos\theta)\,\tan\varphi\,-\,\frac{\cos^{^2}\theta}{n\,\cos^{^3}\psi}\,(1\,+\,
\tan\varphi)\,\right]\,\overline{AB}\,\,.
\label{eq:geo}
\end{eqnarray}
The plots of  $z_{_{\rm out}}$, $z_{_{\rm out}}- z_{_{\rm Snell}}^{^{\perp}}$,   and  $z_{_{\rm out}} - z_{_{\rm Snell}}$, for $\overline{SD}=0$, are shown in Fig.\,2 for dielectric blocks made of BK7 (a), LASF9 (b), and  Diamond (c). From these plots, it is clear that the
maximal breaking of the transversal symmetry happens  for the incidence angle $\theta=\pi/4$. For such an incidence,
\begin{equation}
\label{eqzsnell2}
\left\{\, z_{_{\rm Snell}}^{^{\perp}}\,,\,z_{_{\rm Snell}}\, \right\}_{_{\theta=\pi/4}} = N\,\left\{\, \sqrt{2}\,\overline{AB}\,,\,\frac{(n^{\2}-1)\,\overline{BC} + \sqrt{2}\,n^{\2}\,\overline{AB}}{2\,n ^{^{2}}-1}\, \right\}\,\,,
\end{equation}
and consequently
\begin{eqnarray}
z_{_{{\rm Snell},\pi/4}} -\,\, z_{_{{\rm Snell},\pi/4}}^{^{\perp}} & =&  N\,\left[\, \frac{(n^{\2}-1)\,\overline{BC} + \sqrt{2}\,n^{\2}\,\overline{AB}}{2\,n ^{^{2}}-1}\,-\,   \sqrt{2}\,\overline{AB}     \, \right]\,\,,   \nonumber \\
  & = & N\, \frac{n^{\2}-1}{2\,n ^{^{2}}-1}\,\left(\,\overline{BC}-\sqrt{2}\,\,\overline{AB}\,\right) =   \sqrt{2}\,\,\,\frac{1+\sqrt{2\,n^{^{2}}-1}}{2\,n^{^{2}}-1}\, \,N\,\overline{AB}\,\,.
  \label{dz}
\end{eqnarray}
 Observing that  from Eq.\,(\ref{eq:Eout3}), by using $y_{_{{\rm Snell},\pi/4}}=0$, we obtain the following transmission intensity
 \begin{equation*}
\mbox{I}_{_T}^{^{[s,\,p]}}(\boldsymbol{r})_{_{\theta=\pi/4}}= \,\,
\frac{\mbox{w}_{_{0}}^{^{2}}\,\mbox{I}_{_0}\,\mbox{P}^{^{[s,\,p]}}_{_{T,\pi/4}}}{\mbox{w}(z-z_{_{{\rm Snell},\pi/4}}^{^{\perp}})\,\mbox{w}(z-z_{_{{\rm Snell},\pi/4}})}\,\,\exp\left\{\,-\,2\,\left[\,\frac{x^{^{2}}}{\mbox{w}^{^{2}}(z-z_{_{{\rm Snell},\pi/4}}^{^{\perp}})} + \frac{y^{^{2}}}{\mbox{w}^{^{2}}(z-z_{_{{\rm Snell},\pi/4}})} \,\right]\right\}
 \end{equation*}
Eq.\,(\ref{dz}) clearly implies that the beam spot size is not more symmetric in the transversal planes $xz$ and $yz$. To estimate the breaking of symmetry, we can calculate
the transversal spot size difference at the exit point $z=z_{_{{\rm out},\pi/4}}=\overline{SD}+N\,\overline{BC}$,
\begin{eqnarray}
\Delta\mbox{w}^{^{2}}_{_{\pi/4}}     &=& \mbox{w}^{^{2}}(z_{_{{\rm out},\pi/4}}-z_{_{{\rm Snell},\pi/4}}^{^{\perp}}) -  \mbox{w}^{^{2}}(z_{_{{\rm out},\pi/4}}-z_{_{{\rm Snell},\pi/4}}) \nonumber \\
 & = &
\frac{4}{(\,k\,\mbox{w}_{_{0}})^{^{\,2}}}\,\left(\, z_{_{{\rm Snell},\pi/4}} - z_{_{{\rm Snell},\pi/4}}^{^{\perp}}   \,\right) \,\left(\,2\,z_{_{{\rm out},\pi/4}} - z_{_{{\rm Snell},\pi/4}} - z_{_{{\rm Snell},\pi/4}}^{^{\perp}}\,\right) \nonumber
 \\
 & = &\frac{4}{(\,k\,\mbox{w}_{_{0}})^{^{\,2}}} \,\,
 \frac{n^{\2}-1}{2\,n ^{^{2}}-1}\,\,N \left(\,\overline{BC}-\sqrt{2}\,\,\overline{AB}\,\right)
 \left[\, 2\,\overline{SD}\, +    \frac{3\,n^{\2}-1}{2\,n ^{^{2}}-1}\,\,N \left(\,\overline{BC}-\sqrt{2}\,\,\overline{AB}\,\right)      \,\right]\,\,.
\end{eqnarray}
 For convenience of presentation and in view of possible experimental implementations, we explicitly calculate the quantity  $\langle \Delta\mbox{w}\rangle_{_{\pi/4}}=\sqrt{\Delta\mbox{w}^{^{2}}_{_{\pi/4}}}$ in the case of a laser source very close to the first left interface, $\overline{SD}\approx0$, and for BK7,
 LASF9, and Diamond dielectric blocks. From the previous equation, we get
\begin{equation}
\langle\Delta\mbox{w}\rangle_{_{\pi/4,\,\{{\rm BK7,\,LASF9,\,Diamond}\}}}\approx N\,\left\{\,  4.86\,,\,3.25\,,\,2.09  \,\right\}\,\,
\frac{\overline{AB}}{k\,\mbox{w}_{_{0}}}  \,\,.
\end{equation}
It should be noted that in optical experiments $\overline{AB}$ is of the order of cm, and consequently, for a red He-Ne laser ($\lambda=633$\,nm), we find, for beam waist of $100\,\mu{\rm m}$ and $1\,{\rm mm}$,
the following symmetry breaking
\begin{eqnarray}
\langle\Delta\mbox{w}\rangle_{_{\pi/4,\,\{{\rm BK7,\,LASF9,\,Diamond}\}}}^{^{[\,100\,\mu\,] }}&\approx& N\,\left\{\,  4.86\,,\,3.25\,,\,2.09  \,\right\}\,\,\times\,\,10^{^{-1}}\,\mbox{w}_{_{0}} \nonumber \\
 & \approx &   N\,\left\{\,  4.86\,,\,3.25\,,\,2.09  \,\right\}\,\,\times\,\,10\,\mu{\rm m}
\end{eqnarray}
and
\begin{eqnarray}
\langle\Delta\mbox{w}\rangle_{_{\pi/4,\,\{{\rm BK7,\,LASF9,\,Diamond}\}}}^{^{[\,1\,{\rm mm}\,] }}&\approx& N\,\left\{\,  4.86\,,\,3.25\,,\,2.09  \,\right\}\,\,\times\,\,10^{^{-3}}  \, \mbox{w}_{_{0}} \nonumber \\
 & \approx &   N\,\left\{\,  4.86\,,\,3.25\,,\,2.09  \,\right\}\,\,\times\,\,1\,\mu{\rm m}         \,\,.
\end{eqnarray}
This suggests experimental proposals in which the beam waist is of $100\,\mu{\rm m}$.

We conclude this section observing that for incidence perpendicular to the first left interface, i.e.  $\theta=0$, we find
\begin{equation}
 z_{_{{\rm Snell},0}}^{^{\perp}}=\,z_{_{{\rm Snell},0}} = \, N\,\left(\,1-\,\frac{2}{n}\,\right)\,\overline{AB}
\end{equation}
and, consequently, we recover the transversal symmetry, see Fig.\,2.

\section*{\large \color{\ct} IV. MODIFYING THE AXIAL SPREADING}

In the previous section, we have shown one of the most important consequence of the second order term expansion of the optical Snell phase on the outgoing beam propagating through dielectric blocks. Besides
the breaking of the transversal symmetry, the second order term expansion also leads to a modification of the axial spreading. In order to understand this additional phenomenon, let us consider the case of normal incidence.
From Eqs.\,(\ref{eqzout}) and (\ref{eqzsnell1}), for $\theta=0$,  we have
\begin{equation}
\left[\,\begin{array}{c}
 z_{_{\rm out}}\\
 z_{_{\rm out}} -  z_{_{\rm Snell}}^{^{\perp}}\\
z_{_{\rm out}} -  z_{_{\rm Snell}}
\end{array}
  \,\right]_{_{\theta=0}} = \overline{SD} \,\,+\,\,\left[\,\begin{array}{c}
 1\\
 2\,/\,n\\
2\,/\,n
\end{array}
  \,\right]\,\,N\,\,\overline{AB}\,\,.
\end{equation}
From these analytical formulas, we immediately conclude that the outgoing beam is a symmetric beam which for $n>2$ is characterized by a {\em reduced} axial spreading if compared with the axial spreading of a free gaussian beam,
\[  \mbox{w}_{_{\theta=0}}(z_{_{\rm out}} -  z_{_{\rm Snell}}^{^{\perp}}) = \mbox{w}_{_{\theta=0}}(z_{_{\rm out}} -  z_{_{\rm Snell}}) < \mbox{w}(z_{_{\rm out}})  \,\,.  \]
This means that the gaussian beam propagating through a dielectric block of refractive index greater than 2\, experiences a focalization-like effect. In the case of  BK7, LASF9, and Diamond blocks, we find
\[
\left[\,\begin{array}{c}
 z_{_{\rm out}}\\
 z_{_{\rm out}} -  z_{_{\rm Snell}}^{^{\perp}}\\
z_{_{\rm out}} -  z_{_{\rm Snell}}
\end{array}
  \,\right]_{_{0,\, \,\{{\rm BK7,\,LASF9,\,Diamond}\}  }}  \approx
  \overline{SD} \,\,+\,\,
  \left\{\,
  \left[\,\begin{array}{c}1 \\1.32\\1.32\end{array} \,\right]
  \,,\,
    \left[\,\begin{array}{c}1  \\1.08\\1.08\end{array} \,\right]
    \,,\,
      \left[\,\begin{array}{c}1  \\0.83\\0.83\end{array} \,\right]
  \,\right\}
  \,\,N\,\,\overline{AB}\,\,.
\]
For incidence at $\theta=\pi/4$, independently of the refractive index value, we always found a focalization-like effect. Indeed,  we have
\begin{eqnarray}
\left[\,\begin{array}{c}
 z_{_{\rm out}}\\
 z_{_{\rm out}} -  z_{_{\rm Snell}}^{^{\perp}}\\
z_{_{\rm out}} -  z_{_{\rm Snell}}
\end{array}
  \,\right]_{_{\theta=\pi/4}} &=& \overline{SD} \,\,+\,\, N\, \left[\,\begin{array}{c}
 \overline{BC}\\
 \overline{BC} - \sqrt{2}\,\overline{AB}\\
n^{^{2}}\left(\, \overline{BC} - \sqrt{2}\,\,\overline{AB}\,\right)\,/\,(2\,n^{^{2}}-1)
\end{array}
  \,\right]\nonumber \\
   & = & \overline{SD} \,\,+\,\, \frac{\sqrt{2}}{n^{^2}-1}\, \left[\,\begin{array}{c}
 n^{^{2}}+\sqrt{2\,n^{^{2}}-1}  \\
  1+\sqrt{2\,n^{^{2}}-1}\\
n^{^{2}}  \, (\,1+\sqrt{2\,n^{^{2}}-1}\,)    \,/\,(2\,n^{^{2}}-1)
\end{array}
  \,\right]\,\,N\,\,\overline{AB}\,\,.
\end{eqnarray}
For BK7, LASF9, and Diamond blocks, the previous equations reduce to
\[
\left[\,\begin{array}{c}
 z_{_{\rm out}}\\
 z_{_{\rm out}} -  z_{_{\rm Snell}}^{^{\perp}}\\
z_{_{\rm out}} -  z_{_{\rm Snell}}
\end{array}
  \,\right]_{_{\pi/4,\, \,\{{\rm BK7,\,LASF9,\,Diamond}\}  }}  \approx
  \overline{SD} \,\,+\,\,
  \left\{\,
  \left[\,\begin{array}{c}4.57  \\3.16\\2.02\end{array} \,\right]
  \,,\,
    \left[\,\begin{array}{c}3.42  \\2.01\\1.18\end{array} \,\right]
    \,,\,
      \left[\,\begin{array}{c}2.66  \\1.25\\0.68\end{array} \,\right]
  \,\right\}
  \,\,N\,\,\overline{AB}\,\,.
\]
In order to quantify the axial spreading modifications, for $\overline{SD}\approx 0$ (laser source close to the first left interface) we compare the maximum of the  normalized transmitted intensity at the axial exit point [\,$\mbox{I}_{_{\rm T}}^{^{[s,\,p]}}(0,y_{_{\rm Snell}},z_{_{\rm out}})\,/\,\mbox{P}^{^{[s,\,p]}}_{_{T}}$\,] with that one of a free gaussian beam propagating from the source $S$ to  $z=z_{_{\rm out}}$  [\,$\mbox{I}(0,0,z_{_{\rm out}})$\,],
\begin{equation}
\mathcal{M}{\rm ax} =\, \frac{\mbox{I}_{_{\rm T}}^{^{[s,\,p]}}(0,y_{_{\rm Snell}},z_{_{\rm out}})}{\mbox{P}^{^{[s,\,p]}}_{_{T}}\,\mbox{I}(0,0,z_{_{\rm out}})} = \frac{\mbox{w}^{\2}(z_{_{\rm out}})}{\mbox{w}(z_{_{\rm out}}\,-\,z_{_{\rm Snell}}^{^{\perp}})\,\mbox{w}(z_{_{\rm out}}\,-\,z_{_{\rm Snell}})}\,\,.
\label{eqmax}
\end{equation}
This function is plotted in Fig.\,3 for normal incidence and in Fig.\,4 for $\theta=\pi/4$. From Fig.\,3(a), it is clear the inversion point for the refractive index $n=2$.  The opposite behavior between BK7/LASF9 and Diamond blocks is also evident in Fig.\,3(b) where  the Diamond curve is greater that one (focalization-like effect). The plots have been done for
\[ k\,\mbox{w}_{_{0}}=10^{^{3}}\,\,\,\,\,\,\,\mbox{and}\,\,\,\,\,\,\,
\overline{AB}=10^{^{2}}\mbox{w}_{_{0}}\,\,.    \]
This represents a good approximation  for experiments in which  He-Ne laser ($\lambda=633\,{\rm nm}$) of beam waist $\mbox{w}_{_{0}}=100\,\mu{\rm m}$, and a block side $\overline{AB}$ of $1\,{\rm cm}$ are used.

It is interesting to observe that the focalization effect  is amplified  by increasing the number ($N$) of
the  dielectric blocks. This can be clearly seen  in Fig.\,3(b) for Diamond blocks and in Fig.\,4 for  BVK7,LASF9, and Diamond blocks.  From these plots, we can also note that the curves which determine the axial spreading behavior  tend to a fixed value for an increasing number of blocks. In this limit,
after simple algebraic manipulations, we get
\begin{equation}
\label{plateau}
\left\{\,\mathcal{M}{\rm ax}_{_{\theta=0}} \,,\, \mathcal{M}{\rm ax}_{_{\theta=\pi/4}}\,\right\}\,\,\,\to\,\,\,
\left\{\,\frac{n^{^{2}}}{4}\,,\,\frac{(2\,n^{^{2}}-1)\, \left(\,n^{^{2}}+\sqrt{2\,n^{^{2}}-1}\,\right)^{^{2}}}{n^{^{2}}  \,
\left(\, 1+\sqrt{2\,n^{^{2}}-1}\,\right)^{^{2}}}\,\right\}\,\,.
\end{equation}
Note that the number of blocks $N$ does not interfere in the normalization $\mbox{P}^{^{[s,\,p]}}_{_{T}}$. This is due the fact that  for BK7, LASF9, and Diamond blocks the critical angles are
\[\theta_{{c,\, \{{\rm BK7,\,LASF9,\,Diamond}\}}}  =-\,\left\{\,5.6^{^{o}}\,,\,22.9^{^{o}}\,,\,57.7^{^{o}}\,\right\}\,\,,   \]
and consequently, for normal incidence and for incidence at $\theta=\pi/4$, we are in the situation of total internal reflection.  Independently  of the block numbers, we have
\[  \mbox{P}^{^{[s]}}_{_{T,\,0,\,  \{{\rm BK7,\,LASF9,\,Diamond}\}    }}  =\mbox{P}^{^{[p]}}_{_{T,\,0,\,  \{{\rm BK7,\,LASF9,\,Diamond}\}    }}
= \left\{\,
0.92
\,,\,
0.83
\,,\,
0.69
\, \right\}
 \]
and
\[  \mbox{P}^{^{[s,\,p]}}_{_{T,\,\pi/4,\,  \{{\rm BK7,\,LASF9,\,Diamond}\}    }}  =
\left\{\,
\left[\,0.82 \,,\, 0.98 \, \right]
\,,\,
\left[\,0.69 \,,\,0.94  \, \right]
\,,\,
\left[\,0.52 \,,\, 0.85 \, \right]
\, \right\}
 \]
The predicted plateau values of Eq.\,(\ref{plateau}), for normal incidence
\[
\mathcal{M}{\rm ax}_{_{0,\,  \{{\rm BK7,\,LASF9,\,Diamond}\}}} =
\left\{\,
0.57
\,,\,
0.85
\,,\,
1.45
\,\right\}\,\,,
\]
and for incidence at $\theta=\pi/4$
\[
\mathcal{M}{\rm ax}_{_{\pi/4,\,  \{{\rm BK7,\,LASF9,\,Diamond}\}}} =
\left\{\,
3.28
\,,\,
4.96
\,,\,
8.30
\,\right\}
\]
are confirmed by the numerical calculations presented in the plots of Figs.\,3 and 4.

\section*{\large \color{\ct} V. CONCLUSIONS AND OUTLOOKS}

The main objective of the investigation presented in this paper  was the detailed analysis of the second order term in the Taylor expansion of the Snell phase. This study was stimulated by the fact that this  geometrical phase, whose first order term in its Taylor expansion has been recently used to to demonstrate the Snell law\cite{AJP}, also contains a second order term which, in particular incidence conditions, could produce experimentally detectable changes in the shape of optical beams which propagated through elongated dielectric blocks.   The possibility to use the Snell phase to obtain in a more simple way the optical path predicted by the ray optics  is surely an important point if we look for an elegant derivation  of the optical geometrical path. Nevertheless, before the discussion presented in this paper, the use of such a phase appears as an elegant auxiliary but not necessary  concept to be included in optical calculations. The results of our investigation show two important effects of the Snell phase on the transmitted optical beam.  The first interesting effect is the breaking of symmetry in the transversal components of the beam. The second  one comes from the reduction of the axial spreading.  Before to conclude this paper we briefly discuss a simple experimental proposal to test our theoretical predictions.

Let us first consider the propagation of a free gaussian beam. As well know, it spreads transversally as
it propagates along the $z$-axis.  At $z=0$, the irradiance profile fallen to $1/e$  for points belonging to $x^{\2}+y^{\2}={\rm w}_{\0}^{\2}/2$. After a free propagation, at a distance $z$ from the source, the laser beam wave front is still symmetric and acquires curvature generating a $1/e$-irradiance profile located at
\begin{equation}
 x^{\2}+y^{\2} = {\rm w}^{^2}(z)/2\,\,,
 \end{equation}
where ${\rm w}(z)= {\rm w}_{_{0}} \sqrt{1 +  (\,\lambda\,z/\pi\,{\rm w}_{_{0}}^{^{2}}\,)^{^{2}}}$. As it has been explicitly shown in this paper, after propagating through an elongated  dielectric block, the transmitted beam  is characterized by the following  $1/e$-irradiance profile
\begin{equation}
 \frac{x^{\2}}{{\rm w}^{^2}(z-z_{_{{\rm Snell}}}^{^{\perp}}  )} \,  +\,
\frac{y^{\2}}{{\rm w}^{^2}(z-z_{_{{\rm Snell}}})} \,=\, \frac{1}{2}\,\,.
\end{equation}
By considering a dielectric structure composed by  $N$ blocks and located at  $d_{_{\rm source}}$ from the source and  $d_{_{\rm cam }}$ from the camera, to estimate the curvature of the free and transmitted beam,
we have to calculate  ${\rm w}(z)$,  ${\rm w}(z-z_{_{{\rm Snell}}}^{^{\perp}})$, and ${\rm w}(z-z_{_{{\rm Snell}}})$ at
\[ z= z_{_{\rm cam}} = d_{_{source}} + D_{_{N\rm blocks}} + \,\,d_{_{\rm cam}}\,\,.\]
Let us consider
\[\lambda=0.633\,\mu{\rm m}\,\,,\,\,\,\,\,{\rm w}_{_0}=100\,\mu{\rm m}\,\,,\,\,\,\,\,\overline{AB}=1\,{\rm cm}\,\,,\,\,\,\,\,  d_{_{source}} +\,\, d_{_{\rm cam}} = 5\, {\rm cm}\,\,,\,\,\,\,\,
n=n_{_{\rm BK7}}
\,\,\,\,\,{\rm and}\,\,\,\,\,N=5\,\,.\]
The previous choices imply
\[\frac{2\,N\,\overline{AB}}{k\,{\rm w}_{_{0}}^{^{2}}}\, =  \frac{2\,(\,d_{_{source}} +\,\, d_{_{\rm cam}})}{k\,{\rm w}_{_{0}}^{^{2}}}\,\approx \,1\,\,\,\,\,\,\,{\rm and}\,\,\,\,\,\,\,
D_{_{\rm block}}=N\,\overline{BC}\, \approx 4.57\,N\,\overline{AB}\,=22.85\,{\rm cm}\,\,.\]
For free propagation, we thus obtain a  $1/e$-irradiance contour at the circle of radius
\begin{equation}
r =\frac{{\rm w}(z_{_{cam}})}{\sqrt{2}} =\sqrt{\frac{1+5.57^{^{\,2}}}{2}}\,{\rm w}_{_0}\approx 4\,{\rm w}_{_0}\,\,,
\end{equation}
see Fig.\,5 (left side). Let us now repeat the measurement by interposing between the source and the camera the  BK7 dielectric block, see Fig.\,5 (right side).  For incidence at $\theta=\pi/4$,  after the propagation through the BK7 block, following our theoretical predictions we should find
a {\em focalized} elliptic $1/e$ irradiance profile
\begin{equation}
\begin{array}{ccccccc}
r_{_x} &=&  \displaystyle{\frac{{\rm w}(z_{_{cam}}- z_{_{{\rm Snell}}}^{^{\perp}})}{\sqrt{2}}} &=& \displaystyle{\sqrt{\frac{1+4.16^{^{\,2}}}{2}}\,{\rm w}_{_0}} & \approx & 3\,{\rm w}_{_0}\,\,\\
r_{_y} &=&  \displaystyle{\frac{{\rm w}(z_{_{cam}}-z_{_{{\rm Snell}}})}{\sqrt{2}}}& =&  \displaystyle{\sqrt{\frac{1+3.02^{^{\,2}}}{2}}\,{\rm w}_{_0}} &\approx & \displaystyle{\frac{9}{4}}\,{\rm w}_{_0}\,\,.
\end{array}
\end{equation}
In this case is clear the influence of the Snell phase in the shape modification of the gaussian beam. Compare the left and right of Fig.\,5.

It is also interesting to be observed that the beam waist dimension plays a fundamental role in the experimental implementation. Indeed, if for example we use a gaussian He-Ne laser with beam waist of
$1\,{\rm mm}$ instead of $100\,\mu{\rm m}$, to reproduce the previous breaking of symmetry we need to
modify the following  entries
\[    d_{_{source}} +\,\, d_{_{\rm cam}} = 5\, {\rm m}\,\,
\,\,\,\,\,{\rm and}\,\,\,\,\,\,\,N=500\,\,.\]
This should require an optical laboratory greater than $30$ m. Without changing these parameters,
we find
\begin{equation}
\left\{\,r\,,\,r_{_{x}}\,,\,r_{_{y}}\,\right\}\approx \frac{1}{\sqrt{2}}\,\left[\,1 + \frac{1}{2}\,\left\{\,
\left(\,\frac{5.57}{100}\,\right)^{^{2}}\,,\,
\left(\,\frac{4.16}{100}\,\right)^{^{2}}\,,\,
\left(\,\frac{3.02}{100}\,\right)^{^{2}}
\,\right\}\,\right]\,\,,
\end{equation}
and practically no real deviation from the $1/e$-irradiance profile of the free beam could be seen.

The theoretical study presented in this paper,  if experimentally confirmed, clearly shows the importance to include the Snell phase in describing the propagation of optical beams through  dielectric blocks. In a forthcoming paper, we aim to extend our analysis to  the optical analog of weak measurements\cite{wm,axial}, where the Snell phase could play an important role in describing the correct behavior of the oscillation between  $s$ and $p$ polarized waves.

\vspace*{1cm}

\noindent
\textbf{ACKNOWLEDGEMENTS}\\
The authors  thank  the Capes (M.P.A. and M.L.) and  CNPq (S.D.L.) for the financial support.
They also gratefully thank  Silv\^ania A. Carvalho  and Gabriel G. Maia for useful discussions and an anonymous referee for his comments and suggestions.

\newpage

\begin{figure}[ht]
% 	\centering
	\vspace{-2.2cm}
		\hspace{-0.8cm}
		\includegraphics[height=22cm, width=17cm]{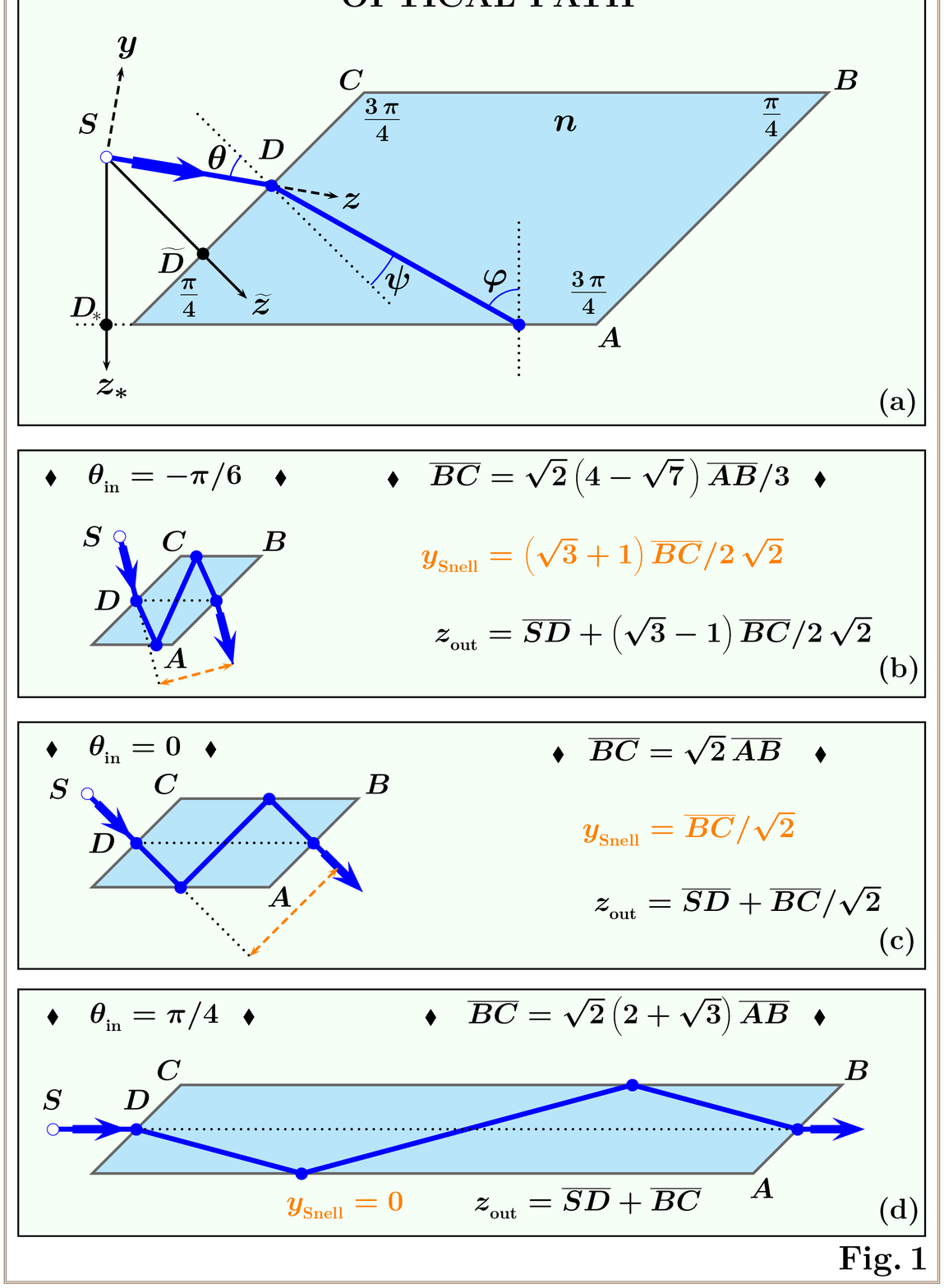}
		\vspace{-1cm}
		\caption{{\bf Dielectric block.} The gaussian beam propagates, along the $z$-axis, from its source up to the first (left) air/dielectric interface. The incidence angle with respect to the normal to the left interface, $\widetilde{z}$-axis, is $\theta$. The angle of incidence at the second (down) dielectric/air  interface is $\varphi$. The condition $\overline{BC}=\sqrt{2}\,\tan\varphi\,\overline{AB}$ is chosen to guarantee an outgoing beam with the same $z_*$ component of the incoming one. This allows to realize an elongated dielectric structure, done by $N$ blocks, in which the light propagation suffer   $2\,N$ internal reflections.  In the figures (b-d), the Snell shift, $y_{_{\rm Snell}}$, and the axial position of the beam at the exit point, $z_{_{\rm out}}$, are given for three incidence angles, $\theta=-\,\pi/6,\,\,0,\,\,\pi/4$, and for a dielectric block of refractive index $n=\sqrt{2}$.}
	\label{fig:fig1}
\end{figure}

\newpage

\begin{figure}[ht]
% 	\centering
	\vspace{-2.2cm}
		\hspace{-0.8cm}
		\includegraphics[height=22cm, width=17cm]{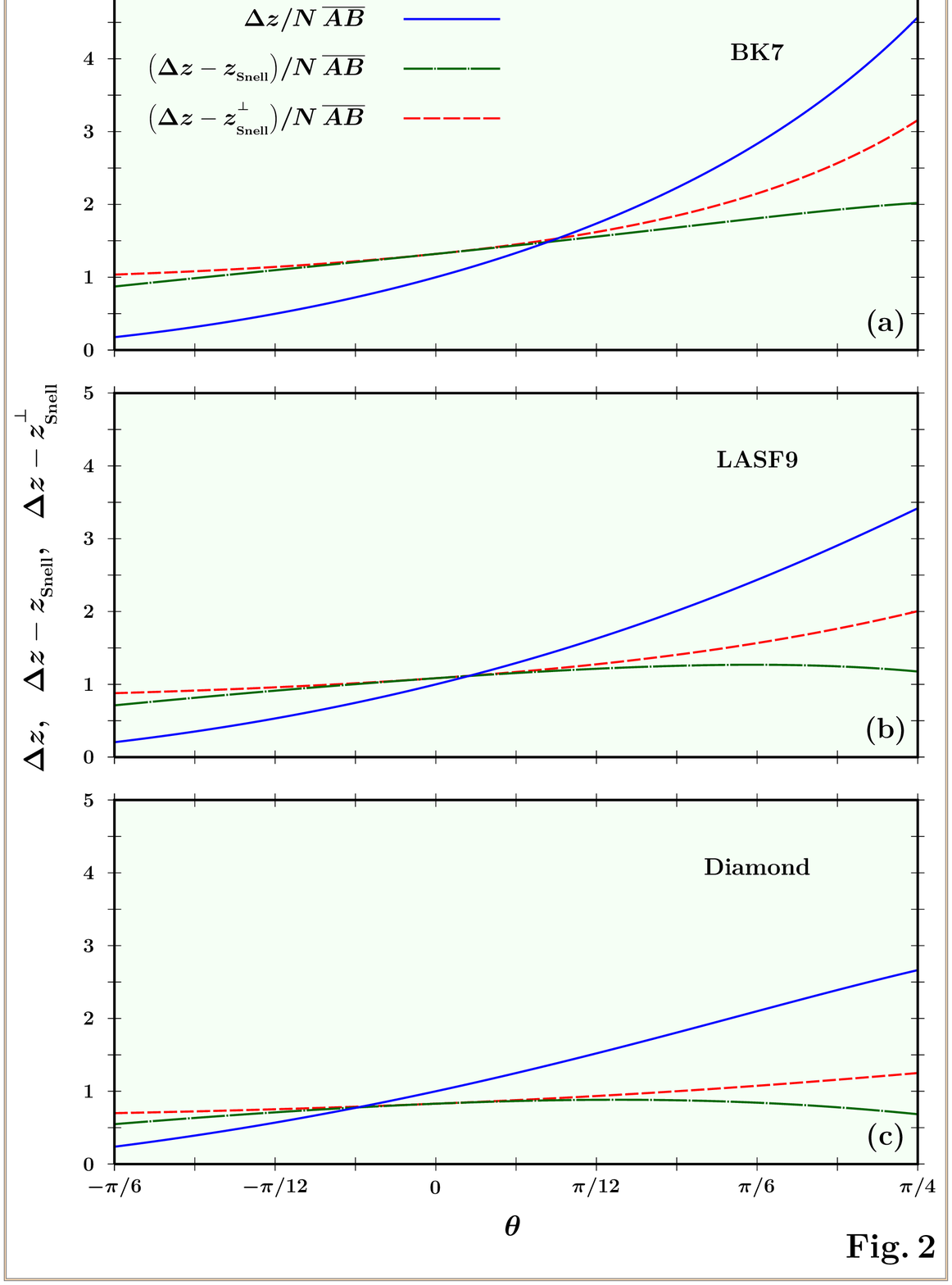}
		\vspace{-1cm}
		\caption{ {\bf Transversal spreading.}  The beam spreading  of a free gaussian beam, which propagates  from its source up to $z= z_{_{\rm out}}$, is compared with the transversal spreading of a beam propagating thorough  an elongated dielectric structure composed by $N$ blocks, $\mbox{w}(z_{_x}^{^{\rm [tra]}})= \mbox{w}(z_{_{\rm out}}- z_{_{\rm Snell}}^{^{\perp}})$ and  $\mbox{w}(z_{_y}^{^{\rm [tra]}})= \mbox{w}(z_{_{\rm out}}- z_{_{\rm Snell}})$.  From the plots is clear the breaking of symmetry fro incidence angle far from the normal incidence. The intersection point between free propagation and dielectric transmission represents the incidence angle for which we begin to see the focalization-like effect.  For a dielectric with refractive index $n=\sqrt{2}$, this intersection point is obtained for incidence at $\theta=0$.      }
	\label{fig:fig2}
\end{figure}

\newpage

\begin{figure}[ht]
% 	\centering
	\vspace{-2.2cm}
		\hspace{-0.8cm}
		\includegraphics[height=22cm, width=17cm]{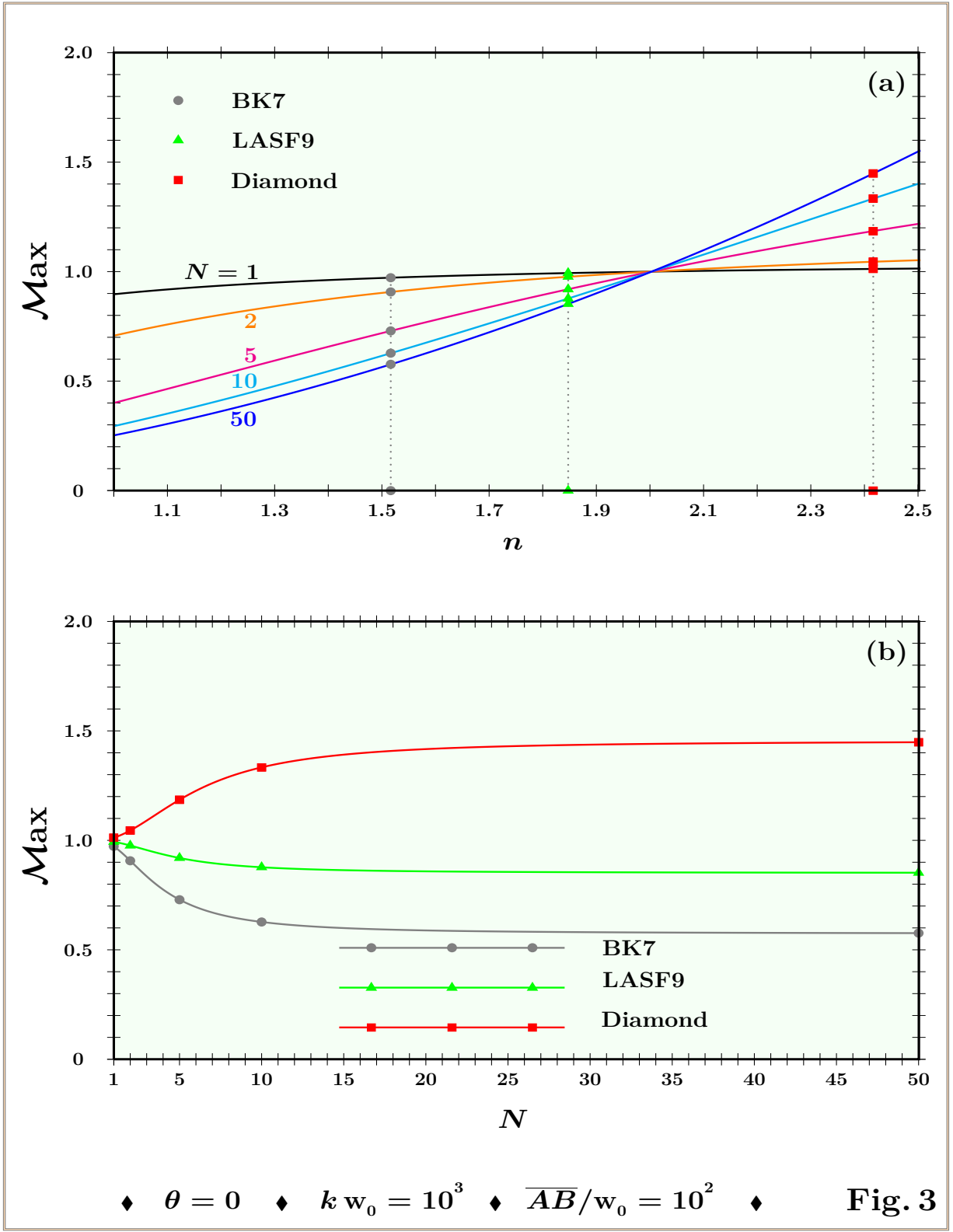}
		\vspace{-1.2cm}
		\caption{{\bf Normal incidence.}  The ratio between the maximum of a free propagating beam  and the maximum of a beam incident at $\theta=0$ and propagating through an elongated dielectric structure composed by $N$ blocks, ${\mathcal M}{\rm ax}$ [see  Eq.\,(\ref{eqmax})], is plotted as a function of the refractive index (a) and of the blocks number (b). For normal incidence, the refractive index $n=\sqrt{2}$ represents the refractive index starting value for which we can see focalization-like effects, see (a). Increasing the number of blocks, we reach a limit value, see (b). Consequently, in view of a possible experimental implementation,   a good choice of number of blocks is represented by $5\leq N\leq 10$. }
	\label{fig:fig3}
\end{figure}

\newpage
\begin{figure}[ht]
% 	\centering
	\vspace{-2.2cm}
		\hspace{-0.8cm}
		\includegraphics[height=22cm, width=17cm]{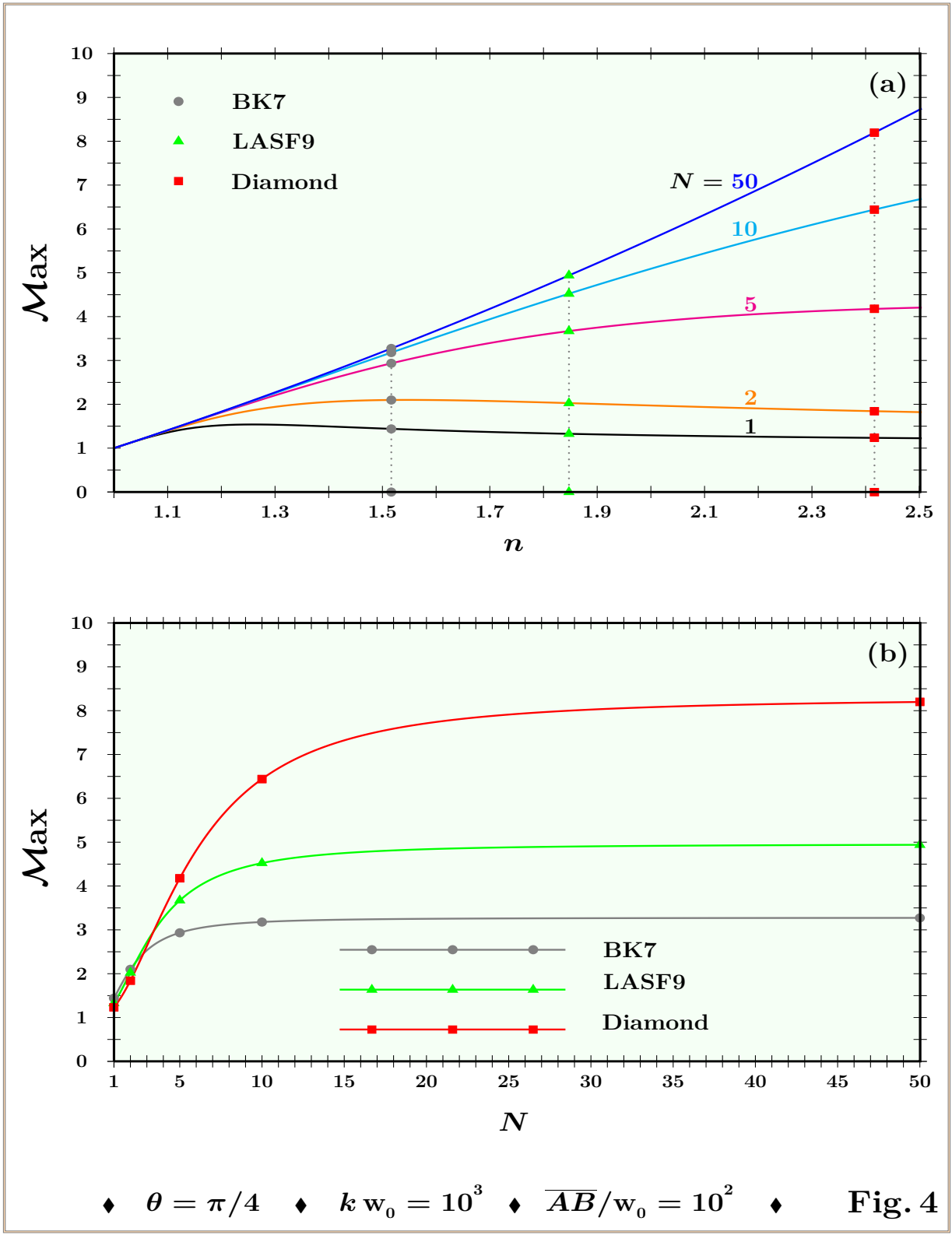}
		\vspace{-1.2cm}
		\caption{{\bf Incidence at $\boldsymbol{\theta=\pi/4}$.}  The ratio between the maximum of a free propagating beam  and the maximum of a beam incident at $\theta=\pi/4$ and propagating through an elongated dielectric structure composed by $N$ blocks, ${\mathcal M}{\rm ax}$ [see  Eq.\,(\ref{eqmax})], is plotted as a function of the refractive index (a) and of the blocks number (b). For such an incidence, independently of the refractive index, we always find a focalization-like effect, ${\mathcal M}{\rm ax}>1$, see (a). Increasing the number of blocks, we newly reach a limit value, see (b). For incidence at $\theta=\pi/4$, the maximal amplification of the  focalization effect for BK7, LASF9, and Diamond  obtained from the plots in (b), i.e.  3.28, 4.96, and 8.30, confirms the analytical results given  in section IV.  }
	\label{fig:fig4}
\end{figure}

\newpage
\begin{figure}[ht]
% 	\centering
	\vspace{-2.2cm}
		\hspace{-0.3cm}
		\includegraphics[height=21cm, width=15.5cm]{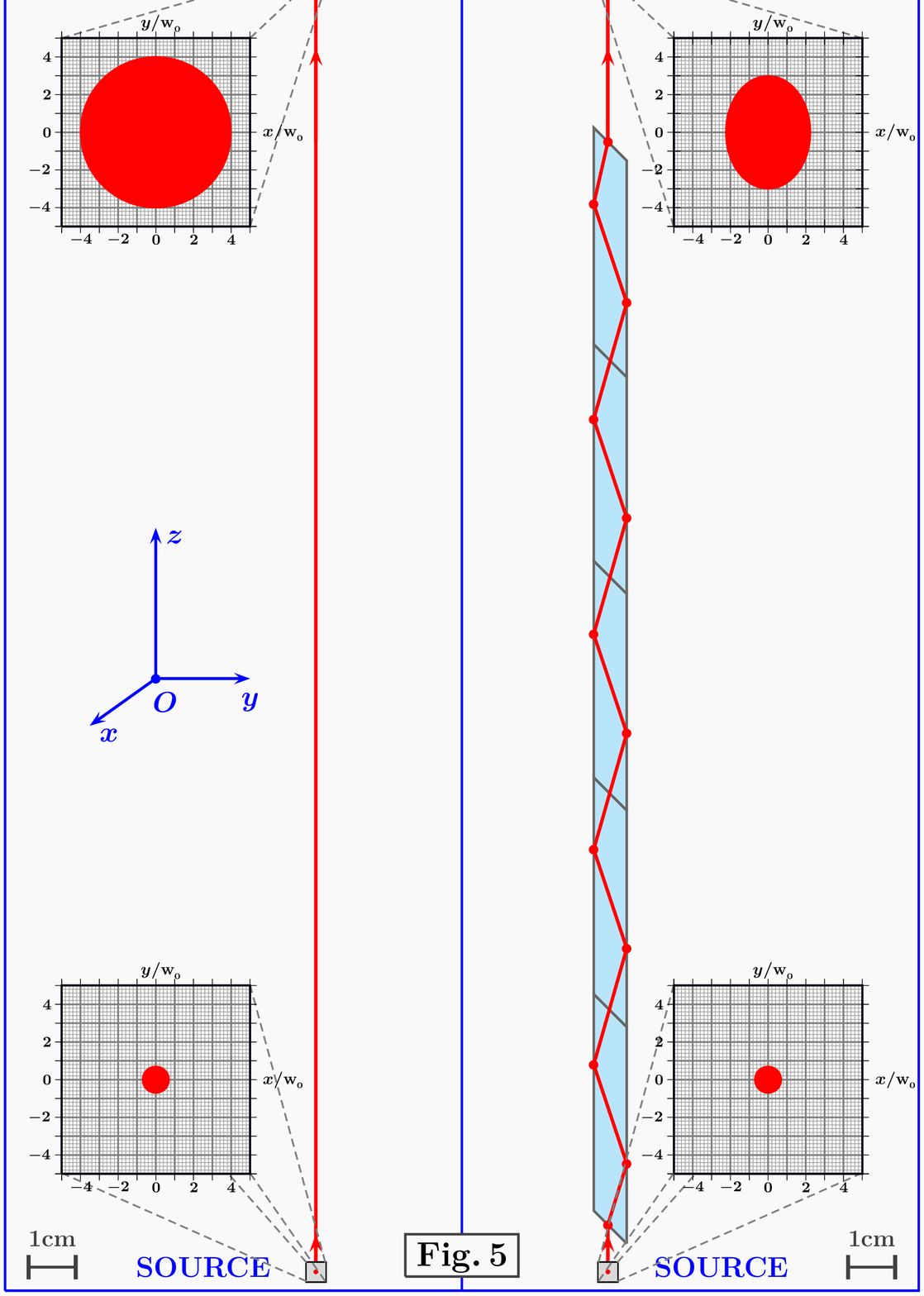}
		\vspace{-0.4cm}
		\caption{{\bf Experimental proposal}. In this figure, we propose an experimental layout to see the breaking oftransversal  symmetry and the focalization-like effect due to the Snell phase.  The propagation of a free gaussian beam, with ${\rm w}_{_{0}}=100\,\mu{\rm m}$ and $\lambda=633\,{\rm nm}$ [see (a)] is compared with the propagation of the same  gaussian beam passing thorough  an elongated dielectric structure composed by 5 BK7 blocks [see (b)]. In this case, each block is characterized by the following sides ratio $\overline{BC}/\overline{AB}=4.574$. At $z=27.87\,{\rm cm} [=1\,{\rm cm} + 5\, \overline{BC} + 5\, {\rm cm}]$, diffraction causes free light waves to spread transversally by increasing the $1/e$ profile radius from the source value ${\rm w}_{_0}/\sqrt{2}$ to the camera value $4\,{\rm w}_{_0}$, see (a). The beam propagating through the BK7 dielectric, at the camera, will present a focalized  elliptic transversal shape
with $r_{_x}=3\,{\rm w}_{_0}$ and  $r_{_y}=9\,{\rm w}_{_0}\,/4$.
}
	\label{fig:fig5}
\end{figure}

\end{document}